\documentclass{article}

\usepackage{amsmath, amssymb, amsthm, amsfonts}
\usepackage{authblk}

\usepackage{graphicx}
\usepackage{subcaption}
\usepackage{enumerate}
\usepackage{epsfig}
\usepackage{cite}

\begin{document}

\title{Reconsideration of feasibility of Hall amplifier}

\author{Abhimanyu Kumar}
\affil{Electrical and Instrumentation Engineering Department, Thapar Institute of Engineering and Technology, Patiala--147004, Punjab, India. Email: akumar2307@outlook.com}

\author{Om Prakash Pandey}
\affil{School of Physics and Material Sciences, Thapar Institute of Engineering and Technology, Patiala--147004, Punjab, India. Email: oppandey@thapar.edu}

\maketitle

\begin{abstract}
In 1955, it was first suggested that Hall effect can be employed for amplification purposes by using semiconductor material with very high mobility. While this idea was limited at that time, yet it was not entirely discarded expecting eventual progress. We revisit this idea and discuss it in the light of current literature. This manuscript kindles this 65 year old amazing idea and views it with modern understanding, which will aid in realizing Hall amplifiers.
\end{abstract}


\leavevmode \\

Ross and Thompson \cite{Ross1955}, in 1955, suggested to design amplifiers employing the Hall effect principle. Their idea relied on using a material with high mobility in the Hall device which might be able to amplify input signal at the expense of applied magnetic field. While this paper appears as the top result when searched with the keyword ``Hall amplifier", interestingly, it has no more than 18 citations since its inception. In fact, there seems to be no scholarly work being done in this direction.

In the same year, Prof. Barlow \cite{Barlow1955} indicated that their idea was amazing but the literature then (and infact even today) was (is) limited to accomplish the task. For a given input in volts, the Hall device produces output in millivolts or lower. This makes it suitable for sensor applications and measurement devices \cite{Ramsden2011}, but obviously not for amplification. In view of this, it is easy to understand why the idea of Hall amplifier was never valued.



Since 1950s, the industry has undergone major changes. We have witnessed the semiconductor revolution which not only caused the extinction of vacuum tubes and several other primitive devices, but has also led to the miniaturization of circuits and development of efficient control apparatus. Several advances have been made in materials employed for Hall devices. Two-dimensional (2D) materials like graphene \cite{Neto2009, Schwierz2010, Novoselov2012} have gained popularity for sensing applications. Table \ref{sensitivitytable} summarizes the Hall sensitivity ($H_c$) values for various materials. 

\begin{table}[ht!]
\caption{Hall sensitivity values for different materials} \label{sensitivitytable}
\centering
\begin{tabular}{c c}
\hline
Material & $H_c$ ($\Omega$/T) \\
\hline
InAs \cite{Behet1998} & 370 \\
InSb \cite{Sandhu2004} & 370 \\
Graphene \cite{Xu2013} & 800 \\
Graphene \cite{Shi2013} & 1200 \\
Graphene \cite{Huang2014} & 2093 \\
Graphene \cite{Chen2015} & 2745 \\
MoS$_2$/h-BN \cite{Joo2017} & 2996 \\
\hline
\end{tabular}
\end{table}

Transition metal chalcogenides and dichalcogenides are also studied \cite{Park2016}. Materials like Al$_{0.20}$Ga$_{0.80}$N/GaN \cite{White2018} and Au/MoSe$_{2}$/Au structure \cite{Abderrahmane2019} with sensitivities $113~\Omega/\text{T}$ and $118.5~\Omega/\text{T}$ are currently emerging in this area. As reported in the table above, higher Hall sensitivity will produce high Hall voltage. 

In addition to appropriate choice of material, it is also required to ensure that Hall material has small thickness. Devices based on semiconductors in their 2D forms offer nano-scale thickness and consume low power. A promising solution would be a material with thin conduction layer and high surface mobility, which are the main requirements for a high magnetic sensitivity \cite{Joo2017}. With the advent of micro-Hall sensors, this requirement of small dimensions is conveniently satisfied.

Even with huge strides made in the field of material science, we know that the requirement of high mobility is not enough for development of Hall amplifiers. One might now think of increasing the magnetic field, but this would require tens of Tesla to even get the output voltage equal to the input. In reality, the output gets saturated at such high magnetism because the linearity range is usually in the order of tens of milli-Tesla. Therefore, in addition to high Hall sensitivity, a low series resistance is necessary. For most materials, the series resistance is of the order of kilo-Ohms which limits their usefulness to sensor applications only \cite{Abderrahmane2019}. We have identified two approaches to this problem.

\begin{figure}[h!]
\begin{center}
\includegraphics[scale=0.8]{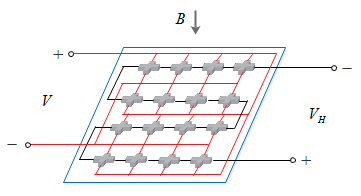}
\caption{Plate composing of several Hall elements} \label{device}
\end{center}
\end{figure}

One approach is to mount several Hall sensors together such that their input ports are in parallel while output ports are in series, as depicted in Figure \ref{device}. The parallel connection on the input side reduces the series resistance, infact it becomes $\frac{R_s}{n}$, where $R_s$ is the series resistance of one Hall element and $n$ is the number of Hall elements. The series connection at output port increases the net Hall voltage given by $V_H = n E_H$, where $E_H$ is the Hall voltage of one Hall element. In case micro-Hall sensors are employed then one can mount hundreds of them on just a unit square-centimeter area. Using a NbFeB permanent magnet (which produces upto 32 mT field) or an electromagnet (for a controllable amplification) to apply magnetic field perpendicular to the sensor plane, one can achieve the desired purpose. 

The second, perhaps more simpler, approach is to use negative resistance elements such as tunnel diode \cite{Degrift1981, Bedrossian1989, Mantegna1994} or Gunn diode \cite{Huang1968, Forster2007, Khalid2014} such that the net series resistance can be decreased. From our experience, we prefer tunnel diodes, majorly due to easiness to use as well as their longevity (as reported in \cite{Esaki2010}). The amplification can probably be better controlled by changing the net series resistance. The decision to use a particular approach is left to the users and the requirement of their specific application. 

For design engineers or experimental physicists, the implementation now is not a difficult task. We believe, in future, the Hall amplifiers will turn into a technology which will be able to compete with the currently available amplifiers.

\end{document}